\documentstyle[12pt,epsfig]{article}

\newcommand{\be}{\begin{equation}}
\newcommand{\ee}{\end{equation}}
\newcommand{\bea}{\begin{eqnarray}}
\newcommand{\eea}{\end{eqnarray}}

\def\nue{{\nu_e}}
\def\anue{{\bar\nu_e}}
\def\numu{{\nu_{\mu}}}
\def\anumu{{\bar\nu_{\mu}}}
\def\nutau{{\nu_{\tau}}}
\def\anutau{{\bar\nu_{\tau}}}
\newcommand{\dm}{\mbox{$\Delta{m}^{2}$~}}

\def\ltap{\ \raisebox{-.4ex}{\rlap{$\sim$}} \raisebox{.4ex}{$<$}\ }

\newcommand{\sig}{\mbox{$\sigma^{2}$}}
\newcommand{\sia}{\mbox{$\psi_{\alpha}$}}

\newcommand{\sidot}{\mbox{$\overline\psi^{\dot\alpha}$}}

\newcommand{\lh}{\mbox{left-handed~}}

\def\sigmn{\sigma^{\mu \nu}}
\def\barsigmn{\bar{\sigma}^{\mu \nu}}
\def\nue{{\nu_e}}
\def\anue{{\bar\nu_e}}
\def\numu{{\nu_{\mu}}}
\def\anumu{{\bar\nu_{\mu}}}
\def\nutau{{\nu_{\tau}}}
\def\anutau{{\bar\nu_{\tau}}}
\def\nul{{\nu_L}}
\def\anul{{\bar\nu_L}}

\begin{document}

\begin{flushright}
SISSA 101/2003/EP\\
SHEP 03-31     \\
hep-ph/0311326
\end{flushright}

\begin{center}
{\Large \bf Electrophobic 
Lorentz invariance violation for neutrinos
and the see-saw mechanism} \vspace{.5in}


Sandhya Choubey$^{1,2}$ and S.F. King$^{3}$
\vskip .5cm
$^1${{\it INFN, Sezione di Trieste, Trieste, Italy}}\\
$^2${{\it Scuola Internazionale Superiore di Studi Avanzati,
I-34014,
Trieste, Italy}}
\vskip 0.3cm

$^3${\it Department of Physics and Astronomy, University of Southampton, \\
Highfield, Southampton S017 1BJ, UK}

\vskip 1in

\end{center}

\begin{abstract}
We show how Lorentz invariance
violation (LIV) can occur for Majorana neutrinos, without inducing
LIV in the charged leptons via radiative
corrections. Such ``electrophobic'' LIV is
due to the Majorana nature of the 
LIV operator together with electric charge conservation.
Being free from the strong constraints coming from the charged lepton sector,
electrophobic LIV
can in principle be as large as current neutrino experiments permit.
On the other hand electrophobic LIV could be naturally small 
if it originates from LIV in some 
singlet ``right-handed neutrino'' sector, and is felt in the physical
left-handed neutrinos via a see-saw mechanism. 
We develop the formalism appropriate
to electrophobic LIV for Majorana neutrinos, and discuss experimental
constraints at current and future neutrino experiments.
\end{abstract}

\newpage

\section{Introduction}
Lorentz and CPT invariance are considered to be amongst the most
sacred symmetries of elementary particle physics. 
However, this very reason should 
motivate us to search for the smallest of hints of their 
possible violation. Indeed, motivated in part by string theory,
there has been some recent interest in the possibility that 
CPT and Lorentz invariance might be violated in nature
\cite{Colladay:1996iz,Coleman:1998ti}.
CPT violation (CPTV) and Lorentz invariance violation (LIV)
\cite{Colladay:1996iz,Coleman:1998ti}, are clearly interesting effects
but subject to strong constraints coming from charged
fermions. However in the neutrino sector, the limits are much
weaker, 
and so one might hope to observe
such non-standard effects in accurate neutrino oscillation
experiments \cite{Barger:2000iv} due to
CPTV terms of the form 
$\bar{\nu}_L^{\alpha}b^\mu_{\alpha \beta}\gamma^\mu
\nu_L^\beta$, where $\nu_L$ represents left-handed neutrinos
labelled by $\alpha , \beta$ and $b^\mu_{\alpha \beta}$ are
CPTV constants. This operator leads to modifications of the
neutrino oscillation formula as discussed in Appendix A.
A detailed discussion on Lorentz and CPT violation in the neutrino sector 
has recently appeared in \cite{km}, where 
the most general Lagrangian for the
neutrinos in the minimal Standard Model extension is presented,
including a catalogue of all CPTV and LIV terms \cite{km}. 

Although of great potential interest from the point of view
of future neutrino experiments, to stand a chance of the effects being
observable, any CPTV and LIV in the neutrino sector
must be effectively screened from the charged lepton
sector, since the strong limits arising from charged leptons
would already preclude any observation in
neutrino oscillation experiments.
Two main requirements of any effective theory of Lorentz and 
CPT violation in the neutrino sector
are therefore: (i) to explain the smallness of Lorentz and 
CPT violation; (ii) to protect the Lorentz and CPT violation 
in the neutrino sector from the 
bounds coming from the charged lepton sector \cite{km}. 

An elegant way of satisfying (i) is to suppose that such effects originate
in the ``right-handed neutrino'' singlet sector,
and are only fed down to the left-handed neutrino sector via the
see-saw mechanism, thereby giving naturally small LIV in the 
left-handed neutrino sector. This possibility is theoretically
attractive since the ``right-handed neutrinos'' 
could represent any singlet sector,
and need not be associated with ordinary quarks and leptons,
except via their Yukawa couplings to left-handed neutrinos. The
fact that CPT violation is associated only with such a singlet
sector could provide a natural explanation for why CPT appears to
be a good symmetry for charged fermions, while being potentially
badly broken in the neutrino sector. 

Although it is possible to satisfy (i) by appeal to the
see-saw mechanism, in some cases it is not possible 
to satisfy (ii) at the same time.
An example of a problematic case was discussed 
in \cite{Mocioiu:2002pz} for the CPT
violating operator $\bar{N}_R^{\alpha}B'^\mu_{\alpha
\beta}\gamma^\mu N_R^\beta$ for the singlet right-handed neutrinos
$N_R$ labelled by $\alpha , \beta$ and the $B'$ are
CPTV constants in the ``right-handed neutrino'' sector. 
The standard see-saw mechanism
leads to naturally suppressed
CPT violation in the left-handed neutrino sector
of the type mentioned above, namely
$\bar{\nu}_L^{\alpha}b^\mu_{\alpha \beta}\gamma^\mu
\nu_L^\beta$, where now $b^\mu=m_{LR}^2B'^\mu/({B'}^2+M_{RR}^2)$ where
$m_{LR}$ is the Dirac neutrino mass, and $M_{RR}$ is the heavy
Majorana mass of the ``right-handed neutrinos''. However
the problem is \cite{Mocioiu:2002pz} that the see-saw mechanism also
generates unacceptably large CPT violation in the charged lepton sector via
one-loop radiative corrections which yield the operator
$\bar{L}_L^{\alpha}{b_{loop}}^\mu_{\alpha \beta}\gamma^\mu
L_L^\beta$, where $L_L=(\nu_L ~e_L)^T$ is the $SU(2)_L$ doublet
that contains the left-handed charged leptons and neutrinos, where
$b_{electron}\sim b_{loop}^\mu\sim 10^{-2}b^\mu$ \cite{Mocioiu:2002pz}.
Since the limit from the electron is
$b_{electron}<10^{-19}$ eV, this implies that in the neutrino
sector $b<10^{-17}$ eV which renders any CPT violation in the
neutrino sector unobservable.\footnote{To be observable, the
coefficient must be of the same order as an observable neutrino
mass splitting $\sqrt{|\Delta m^2|}$.}

\begin{figure}[t]
\centerline{\psfig{figure=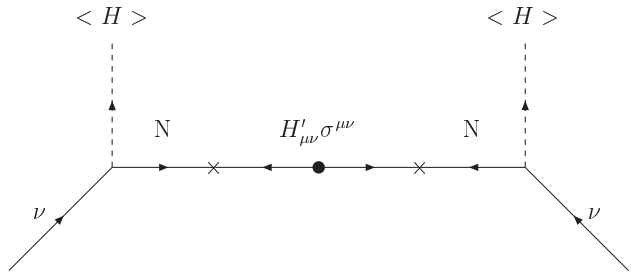,width=4.0in}}
\caption{\label{diag}
{\small Tree level diagram giving rise to electrophobic
see-saw suppressed LIV in the light physical left-handed neutrino 
$\nu$ sector due to a lepton number violating LIV operator in the right-handed
neutrino $N$ sector. The dashed lines represent
(possibly supersymmetric) standard model Higgs with 
vacuum expectation values $<H>$.
The central blob represents insertion of the 
LIV operator in the right-handed neutrino sector, and the crosses
indicate heavy right-handed Majorana masses.}}
\end{figure}

In this paper we show how it is possible
to satisfy both (i) and (ii) at the same
time in a specific example in which the feed-down
into the charged lepton sector is explicitly prevented
by electric charge conservation. 
In this case it is possible to 
have naturally small (but still observable) LIV in the Majorana 
neutrino sector via the see-saw mechanism, without leading
to any LIV in the charged lepton sector via radiative corrections,
to all orders in perturbation theory. 
In order to ensure protection from bounds 
coming from the charged lepton sector the operators should be 
Majorana and lepton number violating.
The essential point is that
such operators, being lepton number violating,
cannot lead to effects in
the charged lepton sector due to electric charge conservation.
As a consequence LIV could be large enough to be observable in the
future neutrino experiments. We refer to such operators as
{\it electrophobic}. We propose that such electrophobic operators
arise exclusively from some heavy ``right-handed neutrino'' singlet sector,
are fed down to physical light left-handed neutrinos via the
see-saw mechanism shown in Fig.~\ref{diag}.
The main purpose of this paper is to develop the
formalism required for phenomenological studies of such 
electrophobic LIV operators, and to briefly
study the experimental constraints at current and planned 
neutrino experiments.

The remainder of the paper is organised as follows.
In section 2 and Appendix B we 
argue that only one electrophobic operator 
exists and show how it can give 
naturally small effects in the left-handed neutrino sector due to the
see-saw mechanism, without inducing any charged lepton
contributions. We also 
derive the equation of motion for the two flavour case.
In section 3 we derive the neutrino oscillation probabilities
in the presence of electrophobic LIV, and 
in section 4 we discuss the experimental constraints
of electrophobic LIV at different experiments. 
Section 5 concludes the paper.
Appendix A contains a derivation of the 
equation of motion and the neutrino survival probability
for the usual CPT violating operator,
$\bar{\nu}_L^{\alpha}b^\mu_{\alpha \beta}\gamma^\mu
\nu_L^\beta$,
while 
Appendix B is dedicated to the details of the electrophobic LIV 
Lagrangian in a two-generation formalism.

\section{Operators, see-saw mechanism and equation of motion}

In this section we shall first
write down LIV operators in some ``right-handed neutrino'' singlet
$N$ sector, arising from some high scale physics, possibly associated
with the scale of heavy Majorana masses $M$. In principle $M$
may be associated with some string scale at which LIV may be manifest
\cite{Colladay:1996iz}, but none of our results depend on this assumption.
In order to be electrophobic
the operator should be Majorana 
in nature, and violate lepton number by $\Delta L=2$.
Therefore we are interested in operators like
$(\overline{N_R^C})_\alpha {\cal O}_{\alpha \beta}
(N_R)_\beta$, where $N_R$ are right-handed neutrinos,
$C$ represents charge conjugation, and
$\cal{O}$ represents the remainder of the operator.
There are only three such non-vanishing Majorana type 
fermion bilinears which break LIV:
\be
{H'}_{\alpha\beta}^{\mu\nu}(\overline{N_R^C})_\alpha \sigma_{\mu\nu}
(N_R)_\beta 
\label{cptR}
\ee
\be
{g'}_{\alpha\beta}^{\mu}(\overline{N_R^C})_\alpha \partial^{\nu}\sigma_{\mu\nu}
(N_R)_\beta;~ {\rm and}~
{g''}_{\alpha\beta}^{\lambda\mu\nu}
(\overline{N_R^C})_\alpha \partial^{\nu}\sigma_{\lambda\mu}
(N_R)_\beta
\label{cptR2}
\ee
where $\mu ,\nu$ denote the Lorentz
indices, $\alpha ,\beta$ are flavor indices
and $H',g', g''$ are LIV constants in the heavy ``right-handed
neutrino'' sector. 
However, since the Majorana singlet that we are concerned with 
are very heavy, the terms in Eq.\ref{cptR2} 
can be dropped in the static limit and the only remaining 
Majorana type LIV term which is relevant is that in Eq.\ref{cptR}.
The see-saw mechanism in Fig.~\ref{diag}
then induces LIV in the left-handed neutrino sector:
\footnote{The term in Eq.\ref{cpt}
looks similar to the magnetic moment operator for Majorana
neutrinos in an electromagnetic field 
${\cal
L}_{EM}=\mu_{\alpha\beta}(\overline{\nu_L^C})_\alpha \sigma_{\mu\nu}
(\nu_L)_\beta F^{\mu\nu}$ \cite{magmom}.
However the magnetic moment operator
respects Lorentz invariance
while the operator in Eq.\ref{cpt} does not. }
\be
{\cal L}_{LIV} \sim 
h_{\alpha\beta}^{\mu\nu}(\overline{\nu_L^C})_\alpha \sigma_{\mu\nu}
(\nu_L)_\beta
\label{cpt}
\ee
where $h^{\mu\nu}=m_{LR}^2{H'}^{\mu\nu}/({H'}^2+M_{RR}^2)$.
As already noted, because of the Majorana nature of the operator, 
LIV cannot be fed down to the
charged lepton sector at any loop order, due to electric charge conservation.
By comparison 
the usual CPT violating operator discussed in Appendix A is not
Majorana, and so the charged lepton sector is not protected
\cite{Mocioiu:2002pz}.

In Appendix B we expand 
${\cal L}_{LIV}$ in a scenario with two neutrino flavors, 
$\nu_\alpha$ and $\nu_\beta$:
\be
{\cal L}_{LIV}=\bigg[(\nu^C_{\alpha R})^\dagger \nu_{\beta L} H_+
-(\nu_{\alpha L})^\dagger \nu_{\beta R}^C H_-
+(\nu_{\beta L})^\dagger \nu_{\alpha R}^C H_- 
- (\nu_{\beta R}^C)^\dagger \nu_{\alpha L}H_+
\bigg ]
\ee
Therefore, in presence of the LIV operator that we consider, 
it is possible for a neutrino of a flavor $\alpha$ to transform 
into an antineutrino of another flavor $\beta$,
as the neutrino beam propagates. This would give 
rise to neutrino-antineutrino 
oscillations between different flavors due to LIV. 
However, transitions between 
neutrino and antineutrino of the same flavor is strictly forbidden, 
since CPT is conserved.

The equation of motion for the two 
neutrino flavour case, including both the mass terms 
and the LIV terms, then follows as, 
\be
i\frac{d}{dt}\pmatrix{\nu_{\alpha L} \cr 
\bar\nu_{\alpha R} \cr \nu_{\beta L} \cr \bar\nu_{\beta R} \cr}
=
\pmatrix{-\delta \cos2\theta & 0 & \delta \sin2\theta & H_- \cr
0 & -\delta \cos2\theta & -H_+ & \delta \sin2\theta \cr
\delta  \sin2\theta & -H_- & \delta  \cos2\theta & 0 \cr
H_+ & \delta  \sin2\theta & 0 & \delta  \cos2\theta \cr}
\pmatrix{\nu_{\alpha L} \cr 
\bar\nu_{\alpha R} \cr \nu_{\beta L} \cr \bar\nu_{\beta R} \cr}
\label{eqofmotliv}
\ee
where $\delta = \Delta m^2/4E$, $\Delta m^2=m_2^2-m_1^2$ is the 
mass squared difference of the neutrinos and 
$E$ is the energy of the neutrino beam.
The equation of motion for Majorana neutrinos with 
non-zero transition magnetic moment 
in the presence of a magnetic field
also has a similar form \cite{rsfp}. 

\section{Neutrino oscillation probabilities}

In this section we look for the 
neutrino transition and survival probabilities in the presence of 
electrophobic LIV interactions. 
The neutrino mass matrix in the 
flavor basis can be written in vacuum as
\be
M_F=\pmatrix
{-\delta \cos2\theta & 0
& \delta \sin2\theta & H \cr
0 & -\delta  \cos2\theta 
& -H & \delta \sin2\theta \cr
\delta \sin2\theta & -H & \delta  \cos2\theta & 0 \cr
H & \delta \sin2\theta & 0 & \delta \cos2\theta \cr}
\label{mf}
\ee
where $H$ is the extra element due to LIV interaction. 
In the above we have expressed 
$H_\pm=He^{\pm i \phi}$
and have assumed $\phi=0$. 
We will show later that $\phi$ is related to the change in the 
neutrino oscillation probabilities
with the direction of the propagation 
of the neutrino and is therefore an important parameter \cite{km}. 
However for the sake of simplicity and to get an approximate 
idea about the constraint on the electrophobic LIV term from 
current and planned experiments, we choose to put $\phi=0$ in this 
and the next section. The case $\phi \neq 0$ will be considered
in section 5.
The mass matrix in Eq. \ref{mf} 
can be diagonalised and the eigenvalues are
\be
\lambda_1 = -\sqrt{ \delta^2 + H^2};~~
\lambda_2 = -\sqrt{ \delta^2 + H^2};~~
\lambda_3 = \sqrt{ \delta^2 + H^2};~~
\lambda_4 = \sqrt{ \delta^2 + H^2}
\ee
The corresponding mixing matrix $\tilde{U}$ 
in the presence of the LIV term is defined as 
\be M_F = \tilde{U}^\dagger diag(\lambda_1,\lambda_2,\lambda_3,\lambda_4)
\tilde{U}\ee and is given by
\be
\tilde{U} = \frac{1}{\sqrt{2}}\pmatrix{\delta \sin2\theta/D_- &
H/D_- & \delta \sin2\theta/D_+ & H/D_+ \cr
-H/D_- & \delta\sin2\theta /D_- & - H/D_+ 
& \delta\sin2\theta /D_+ \cr
-D_-/\sqrt{ \delta^2 + H^2} & 0 & D_+/\sqrt{ \delta^2 + H^2} & 0 \cr
0 & - D_-/\sqrt{ \delta^2 + H^2} & 0 & D_+ /\sqrt{ \delta^2 +H ^2}
}
\label{mm}
\ee
where
\be
D_\pm = \sqrt{ \delta^2 +H ^2}\bigg(1\pm \frac{\delta \cos2\theta}
{\sqrt{ \delta^2 + H^2}}\bigg)^{1/2}
\ee
One can check that when $H=0$, the mixing 
matrix $\tilde{U}$ reduces to the vacuum mixing matrix in the standard case
and there is no mixing between the neutrino and antineutrino states.

The general transition probability of a given flavor $\nu_\alpha$ to 
a flavor $\nu_\beta$ is given by
\be
P_{\alpha\beta}=\delta_{\alpha\beta} - 4\sum_{j>i}
Re(\tilde{U}_{\alpha i} \tilde{U}_{\beta i}^* 
\tilde{U}_{\alpha j}^* \tilde{U}_{\beta j})
\sin^2\bigg(\frac{\Delta\lambda_{ij} L}{2}\bigg)
+2\sum_{j>i} Im(\tilde{U}_{\alpha i} \tilde{U}_{\beta i}^
* \tilde{U}_{\alpha j}^* \tilde{U}_{\beta j})
\sin\bigg(\Delta \lambda_{ij} L\bigg)
\label{probab}
\ee
where $L$ is the distance 
traveled and $\Delta \lambda_{ij} = \lambda_i - \lambda_j$. We 
will assume that the mixing matrix $\tilde{U}$ is real so that 
the last term in Eq.(\ref{probab}) vanishes. Next we note that
$\Delta \lambda_{12} = \Delta \lambda_{34} = 0$ and 
$\Delta \lambda_{13}=\Delta \lambda_{14}=\Delta \lambda_{23}
=\Delta \lambda_{24} = 2\sqrt{ \delta^2 + H^2}$. Thus 
the expression for the probability in the two-generation 
limit that we consider here reduces to
\be
P_{\alpha\beta}=\delta_{\alpha\beta} - 4
\bigg[\tilde{U}_{\alpha 1} \tilde{U}_{\beta 1} + 
\tilde{U}_{\alpha 2} \tilde{U}_{\beta 2}\bigg]
\bigg [ \delta_{\alpha\beta} - (\tilde{U}_{\alpha 1} \tilde{U}_{\beta 1} + 
\tilde{U}_{\alpha 2} \tilde{U}_{\beta 2})\bigg]
\sin^2\bigg(\frac{\sqrt{ (\Delta m^2)^2 + (4HE)^2} ~L}{4E}\bigg)
\label{prob}
\ee
We can now use Eq.(\ref{mm}) and (\ref{prob}) to get 
\bea
P_{\alpha\alpha}&=&P_{\bar \alpha \bar \alpha}
\nonumber \\
&=&
1 - \frac{(4HE)^2 + (\Delta m^2)^2\sin^22\theta}{(4HE)^2+(\Delta m^2)^2 }
\sin^2\bigg(\frac{\sqrt{ (\Delta m^2)^2 + (4HE)^2} ~L}{4E}\bigg)
\label{surv}
\eea
\bea
P_{\alpha \beta}
&=&
\frac{(\Delta m^2)^2\sin^22\theta}{(4HE)^2+(\Delta m^2)^2 }
\sin^2\bigg(\frac{\sqrt{ (\Delta m^2)^2 + (4HE)^2} ~L}{4E}\bigg)
\label{trans1}
\eea
\bea
P_{\alpha \bar \beta}
&=&
\frac{(4HE)^2}{(4HE)^2+(\Delta m^2)^2 }
\sin^2\bigg(\frac{\sqrt{ (\Delta m^2)^2 + (4HE)^2} ~L}{4E}\bigg)
\label{trans2}
\eea
while $P_{\alpha \bar \alpha} =0$ identically, since CPT 
is conserved. 
It is again trivial to see that for $H<<\Delta m^2$ 
or $H=0$, the expressions for the probability reduces to 
the vacuum oscillation probabilities. On the other hand if 
$H \gg \dm$ then,
\bea
P_{\alpha\alpha}=
1 -  \sin^2\bigg(H ~L \bigg)
\label{surv0}
\eea
\bea
P_{\alpha \beta} = 0
\eea
\bea
P_{\alpha \bar \beta} = \sin^2\bigg(H ~L\bigg)
\label{trans0}
\eea
This implies {\it maximal} conversions of the neutrino state 
$\nu_\alpha$ to the antineutrino state  
$\bar\nu_\beta$. But more importantly we note that the 
oscillations are {\it energy independent}. For the case of the 
usual CPT violating operator, 
$\anul^\alpha b_{\alpha \beta}^\mu
\gamma_\mu\nul^\beta $,
the survival probability given by Eq.\ref{prcpt}
in the {\it pure} CPT limit, also
has the same form and is energy independent \cite{Barger:2000iv}.
Another case 
where the survival probability for the atmospheric neutrinos 
have the form given by Eq.\ref{surv0} was considered in \cite{lisire},
again for a CPT violating theory.
The form of the probability considered 
for LIV in \cite{lisire} had a different 
energy dependence. 
The expressions for the survival probability that we derive here, are valid 
for a theory which does not respect Lorentz invariance, however the
CPT symmetry is conserved. We derive the expressions for the 
probabilities in the massless neutrino limit, as well as 
for the case where both neutrino mass and LIV play a role in 
oscillations.

The expressions Eq.\ref{surv}-\ref{trans0} 
are also valid for a theory with 
neutrino transition magnetic moment, in which both Lorentz invariance 
and CPT are conserved. However note 
that $H$ corresponding to 
neutrino magnetic moment is non-zero only in the presence of 
an electromagnetic field. Therefore the case for magnetic moment 
is important only in the presence of an external magnetic field.
Stringent bounds on the neutrino transition magnetic moment $\mu$ can be 
placed from solar and astrophysical data \cite{rsfp}. However 
the LIV term, if non-zero, 
is always present, irrespective of any other condition.

\section{Experimental constraints}

Bounds on electrophobic LIV, parametrised for example by 
the coefficient $H$ discussed in the previous section, 
can be obtained from disappearance 
experiments using Eq.(\ref{surv}), and from appearance 
experiments using Eqs.(\ref{trans1}) and/or (\ref{trans2})
\footnote{We reiterate that the bounds obtained using 
Eq.(\ref{surv}), (\ref{trans1}) and (\ref{trans2}) are approximate
due to the neglect of the direction dependence of the oscillation 
probabilities.}.
While the only appearance experiment with a positive 
signal is LSND, among the most prominent disappearance experiments are 
the solar neutrino experiments, the atmospheric neutrino experiments and
the reactor neutrino experiments, including KamLAND and 
CHOOZ/Palo Verde. 

{\bf Constraints from CHOOZ/Palo Verde:}
The CHOOZ and Palo Verde short baseline reactor experiments are 
consistent with no observed oscillation of $\anue$ at 
baseline $L\sim 1$ km \cite{chooz}. This non-observation of any 
oscillations can be used to constrain $H_{e \bar\beta} \ltap 10^{-19}$ GeV, 
$H_{e \bar\beta}~(=H_{\bar e \beta}$ due to CPT invariance) 
is the LIV coeffecient responsible for 
$\anue(\nue) \rightarrow \nu_\beta(\nu_{\bar\beta})$ transition.

{\bf Constraints from the KamLAND experiment:}
KamLAND observes the electron antineutrinos produced in nuclear 
reactors from all over Japan and Korea. The first results from 
KamLAND show a deficit of the antineutrino flux and are consistent with 
oscillations \cite{kl} with \dm and mixing given by the Large Mixing Angle 
(LMA) solution of the solar neutrino problem \cite{kldata}. 
KamLAND being a disappearance experiment is 
insensitive to whether the $\anue$ oscillate into $\numu$ 
due to mass and mixing or $\anumu$ due to LIV. 
Even though the current KamLAND data, 
has a strong evidence for suppression 
of the incident antineutrino flux, the evidence for energy distortion 
of the resultant spectrum is not very strong -- no distortion of the 
spectrum is allowed at the 53\% C.L. \cite{kl}. Therefore the LIV 
driven oscillations can explain the KamLAND data with 
$H_{e \bar\beta} \sim 7.2 \times 10^{-22}$ GeV. 
Though this LIV solution is not as good as 
oscillations with parameters in the 
LMA region, it is still allowed 
by the first results from the KamLAND experiment.
It could be ruled out 
if the future KamLAND data is consistent 
with spectral distortion.

{\bf Constraints from the atmospheric neutrino data:}
The atmospheric neutrino experiments observe a deficit of the 
$\numu$ and $\anumu$ type neutrinos, while the observed $\nue$ and $\anue$ 
are almost consistent with the atmospheric flux predictions.
The LIV term 
would convert $\numu$($\anumu$) into $\anutau$($\nutau$), 
while flavor oscillations 
convert $\numu$($\anumu$) to $\nutau$($\anutau$).
Since the experiments are insensitive to either $\nutau$ or $\anutau$, 
they will be unable to distinguish between the two cases.
However, since the probability for pure LIV case (cf. Eq.\ref{surv0}) 
is independent of the neutrino energy, it gives the same predicted 
suppression for the sub-GeV, the multi-GeV as well as the upward 
muon data. This is in disagreement with the experimental observations.
Therefore just the LIV term alone fails to explain the data and can 
only exist as a small subdominant effect along with mass driven flavor 
oscillations. Since the downward neutrinos do not show any 
depletion one can use Eq.\ref{surv} to put a limit of 
$H_{\mu \bar\tau} \ltap 10^{-20}$ GeV.
 
{\bf Constraints from the future long baseline experiments:}
Better constraints on the LIV coefficient can be obtained in 
experiments which have longer baselines.
The MINOS experiment \cite{minos} in the US and the 
CERN to Gran Sasso (CNGS) experiments, ICARUS and OPERA \cite{cngs},
have a baseline of about 732 km, though the  
energy of the $\numu$ beam in  
MINOS will be different from the energy of the CERN $\numu$
beam. However, since the LIV driven probability is independent 
of the neutrino energy, all these experiment would be expected 
to constrain
$H_{\mu \bar\beta} \ltap 10^{-22}$ GeV. Among the next generation 
proposed experiments, the JPARC project in Japan \cite{jhf} has a 
shorter baseline of about 300 km only, while the NuMI off-axis 
experiment in the US is expected to have a baseline not very different 
from that in MINOS and CNGS experiments.
The best 
constraints in terrestrial experiments 
would come from the proposed neutrino factory experiments, using 
very high intensity neutrino beams propagating over very large 
distances \cite{nufact}. 
Severe constraints, up to $H_{\mu \bar\beta} \ltap 10^{-23}$ 
GeV could be imposed for baselines of $\sim 10,000$ km.

{\bf Constraints from solar neutrinos:}
Neutrinos coming from the sun,
travel over very long baselines $\sim 1.5\times 10^8$ km.
So one could put stringent constraints on 
$H_{e \bar\beta}$ from the solar neutrino data.
However the situation for solar neutrinos is complicated 
due to the presence of large matter effects in the sun.

{\bf Constraints from supernova neutrinos:}
Supernova are one of 
the largest source of astrophysical neutrinos, releasing 
about $3\times 10^{53}$ ergs of energy in neutrinos.
The neutrinos 
observed from SN1987A,
in the Large Magellanic Cloud, had traveled $\sim 50$ kpc to 
reach the earth.
Neutrinos from a supernova in our own galactic center 
would travel distances $\sim 10$ kpc. These would produce 
large number of events in the terrestrial detectors like 
the Super-Kamiokande. The observed flux and the energy 
distribution of the signal can then be used to 
constrain the LIV coefficient.

{\bf Constraints using the time of flight delay technique:}
Up to now we have been considering the impact on the 
resultant neutrino signal at the detector due to spin-flavor 
oscillations in the presence of the LIV term.
The violation of Lorentz invariance could also change the 
speed of the neutrinos and hence cause delay in their time of flight.
The idea is to 
find the dispersion relation for the neutrinos in the presence 
of LIV and extract their velocity $v=\partial E/\partial p$, where 
$E$ is the energy and $p$ the momentum of the neutrino beam.
Then by comparing the time of flight of the LIV neutrinos, 
with particles conserving Lorentz invariance, one could 
{\it in principle} constrain the LIV coefficient.
The presence of the LIV term in the Lagrangian gives a 
see-saw suppressed correction to the mass term. Therefore
\be
v \approx 1 - \frac{m^2 + m^2_{LIV}}{E^2}
\label{vel1}
\ee
where $m$ is the usual mass of the neutrino concerned and 
$ m^2_{LIV}$ is the LIV correction. The impact of the 
LIV correction 
could be 
important for neutrinos coming from cosmological distances. 
Taking into account the expansion of the universe, the 
LIV part of the mass correction introduces a time delay 
given by \cite{grb}
\be
\Delta t \approx \frac{{\xi^\prime}^2_0}{2}\int_{t_e} ^{t_0} 
\frac{a(t)}{a(t_0)}dt
\label{delt}
\ee
where $a(t)$ is the scale factor of the universe,
$t_e$ is the time when the neutrinos are produced,
$t_0$ is the present time and 
we assume that ${\xi^\prime}^2_0 \ll 1$, where 
\be
{\xi^\prime}^2_0 = \frac{m_{LIV}^2}{E_0}
\ee
$E_0 = E(t_0)$ being the energy of the neutrinos when they are 
observed. In principle, if one could estimate the $\Delta t$, 
the limit on $m_{LIV}$ could be used to obtain the corresponding 
limit on the extent of LIV in the neutrino sector,
although, as discussed in \cite{grb}, making such measurements 
in practice will be a formidable challenge.

\section{Direction dependence and the reference frame}

In this section we show that the oscillation probabilities 
change with the direction of propagation of the neutrino 
in the presence of the electrophobic LIV that we consider in this 
paper.
The equation of motion in the 
flavor basis can be written in vacuum as
\be
i\frac{d}{dt}\pmatrix{\nu_{\alpha L} \cr 
\bar\nu_{\alpha R} \cr \nu_{\beta L} \cr \bar\nu_{\beta R} \cr}
=
\pmatrix{-\delta \cos2\theta & 0 & \delta \sin2\theta & He^{i\phi(t)} \cr
0 & -\delta \cos2\theta & -He^{-i\phi(t)} & \delta \sin2\theta \cr
\delta  \sin2\theta & -He^{i\phi(t)} & \delta  \cos2\theta & 0 \cr
He^{-i\phi(t)} & \delta  \sin2\theta & 0 & \delta  \cos2\theta \cr}
\pmatrix{\nu_{\alpha L} \cr 
\bar\nu_{\alpha R} \cr \nu_{\beta L} \cr \bar\nu_{\beta R} \cr}
\label{eqofmotliv2}
\ee
In the above we have expressed 
$H_\pm=He^{\pm i \phi}$. 
We make a co-ordinate transformation so that 
$\nu_\alpha = S\nu_{\alpha}^\prime$, where 
\be
S=diag(e^{i\phi(t)/2},e^{-i\phi(t)/2},
e^{i\phi(t)/2},e^{-i\phi(t)/2})
\ee
Since $S$ is a diagonal matrix, this transformation does not 
change the oscillation probabilities and we still use the same notation 
for the neutrino flavor states.
However the mass matrix changes to,
\be
i\frac{d}{dt}\pmatrix{\nu_{\alpha L} \cr 
\bar\nu_{\alpha R} \cr \nu_{\beta L} \cr \bar\nu_{\beta R} \cr}
=
\pmatrix{-\delta \cos2\theta + \frac{\dot\phi}{2}
& 0 & \delta \sin2\theta & H \cr
0 & -\delta \cos2\theta -\frac{\dot\phi}{2} & -H & \delta \sin2\theta \cr
\delta  \sin2\theta & -H & \delta  \cos2\theta + \frac{\dot\phi}{2}
& 0 \cr
H & \delta  \sin2\theta & 0 & \delta  \cos2\theta - \frac{\dot\phi}{2}\cr}
\pmatrix{\nu_{\alpha L} \cr 
\bar\nu_{\alpha R} \cr \nu_{\beta L} \cr \bar\nu_{\beta R} \cr}
\label{eqofmotliv3}
\ee
where $\dot\phi \equiv d\phi/dt$. Thus the neutrino mixing in the 
presence of the LIV term we consider, and hence the survival and 
transition probabilities, will depend on $\dot\phi$. One can solve 
Eq. (\ref{eqofmotliv3}) to get the expression for 
the oscillation probabilities in presence 
of LIV. 

It has been stressed in \cite{km} that in the presence of LIV 
interaction terms, one has to specify the 
reference frame in which the experiments are performed. They 
define the ``Sun-centered frame'' ($\{e_1,e_2,e_3\}$) as 
standard reference frame. If we define the reference frame in which 
Eq. (\ref{eqofmotliv3}) is derived with a triad of unit vectors, 
$\hat{p}$, $\hat\epsilon_1$ and $\hat\epsilon_2$, where $\hat{p}$ is a unit vector 
along the direction of propagation of the neutrino and 
$\epsilon_1$ and $\epsilon_2$ are the other two orthonormal vectors, then 
our reference frame is related to the standard frame through the 
unitary transformation \cite{km}:
\be
\pmatrix{\hat{p} \cr \hat\epsilon_1 \cr \hat\epsilon_2} =
\pmatrix{\sin\Theta\cos\Phi & \sin\Theta\sin\Phi & \cos\Theta \cr
\cos\Theta\cos\Phi & \cos\Theta\sin\Phi & -\sin\Theta \cr
-\sin\Phi & \cos\Phi & 0 }
\pmatrix{e_1 \cr e_2 \cr e_3}
\label{transform}
\ee
where $\Theta$ and $\Phi$ are the celestial colatitude and 
longitude of propagation \cite{km}. We note that the 
angles $\Theta$ and $\Phi$ change with the rotation of the earth 
and the propagation of the neutrino. This would make 
$\dot\phi$ non-zero and change the oscillation probability.
One can solve Eq. (\ref{eqofmotliv3}) to get the expressions 
of the mixing in presence of LIV and the oscillation probability 
just as we have done in section 3. Or one could make a co-ordinate 
transformation of the mass matrix given in Eq. (\ref{eqofmotliv3})
to the Sun-centered frame using Eq. (\ref{transform}) and then diagonalise 
it to get the oscillation probability in the Sun-centered frame.

Thus neutrino oscillation probabilities in the presence of the 
electrobhobic LIV that we consider  
depend on the direction 
of the propagation of the neutrino. Therefore the naive bounds 
on the LIV co-efficient that we 
have derived in the previous section would be modified once this 
directional dependence is taken into accout. However for the most general 
case for $H_{\mu\nu}$ this could be quite an involved problem.
A much more detailed discussion on the phenomenology of the 
$H_{\mu\nu}$ Lorentz breaking terms can be found in \cite{km}.

\section{Conclusion}

Both Lorentz and CPT violation are usually subject 
to very strong constraints coming from the charged lepton sector.
Although the limits from neutrino experiments are much weaker,
in some cases 
the Lorentz and CPT violation in the neutrino sector
could be fed into the charged lepton sector at the one loop level,
severely restricting the allowed strength of such effects in
the neutrino sector. In this paper we have explored a class of electrophobic
$\Delta L=2$ lepton number violating 
operators that induce LIV into the Majorana neutrino sector, 
while protecting LIV in the charged lepton sector to all orders
of perturbation theory due to electric charge 
conservation. Among the various 
possible combinations, we find that the operator
$h_{\alpha\beta}^{\mu\nu}(\overline{\nu_L^C})_\alpha \sigma_{\mu\nu}
(\nu_L)_\beta$ appears to be the unique 
candidate. This operator is Lorentz invariance violating, but 
it conserves CPT. To explain the smallness of LIV in the neutrino
sector we have assumed that 
LIV is introduced into a ``right-handed neutrino''
sector at some high scale, possibly close to the string scale,
and feeds down into the left-handed 
sector through the see-saw mechanism,
although our phenomenological results are independent of this assumption.
Independently of this
we have developed the phenomenological formalism of the low energy 
electrophobic operator in the light physical neutrino sector.
We have derived the equation of motion for neutrinos in the presence
of electrophobic LIV. For the 
approximate case, where we neglect the dependence of 
the oscillation probabilities on the direction of the neutrino 
propagation, we have
calculated 
the resulting neutrino survival and transition probabilities,
and briefly discussed the constraints on electrophobic
LIV arising from current and future experiments. We have 
highlighted the importance of the 
direction dependence of the oscillation probability, 
peculiar to the class of LIV terms
considered in this paper.

\vskip 1.3cm

\begin{center}
{\bf Acknowledgement}
\end{center}
{The authors gratefully acknowledge Martin Hirsch and Jose Valle for 
their collaboration during the early stages of this project.
S.C. thanks Werner Rodejohann for many helpful discussions.}

\section*{Appendix A: The Usual CPT Violating Operator}

In this Appendix we derive the equation of motion for the 
previously proposed CPT violating operator \cite{Coleman:1998ti}.
The equation of motion in terms of the flavor states can be
written as \be i\frac{d}{dt}{\pmatrix {\nue \cr \numu \cr}}
=H_F{\pmatrix {\nue \cr \numu \cr}} \label{eqofmot} \ee where
$H_F$ is the Hamiltonian in the flavor basis. In this section we consider 
the usual CPT violating term considered in 
\cite{Coleman:1998ti,Barger:2000iv},
\be \anul^\alpha b_{\alpha \beta}^\mu
\gamma_\mu\nul^\beta \ee
They argue that the only surviving CPT
violating component is $b^0_{\alpha \beta}$ (we may call it $b$ henceforth). It
is a non-diagonal matrix in the flavor basis. This term has a form
similar to the matter potential term when the neutrinos travel
in matter. The Lagrangian in presence of this term has the form
\bea {\cal L}&=& u^\dagger i(\partial_0 - \vec{\sigma}\cdot
\vec{\partial})u + \frac{1}{2}(u^T\sigma_2 M u + u^\dagger
M^\dagger \sigma_2 u^*)+
u^\dagger b^0 u\\
&=&u^\dagger i((\partial_0+b^0) - \vec{\sigma}\cdot
\vec{\partial})u + \frac{1}{2}(u^T\sigma_2 M u + u^\dagger
M^\dagger \sigma_2 u^*) \label{lcpt1} \eea We note that the extra
CPT violating term changes the energy component of the 4-momentum
$p^\mu$. 
The dispersion relation for the
neutrino becomes \be E = \frac{M^2}{2p} + b \label{dispersion} \ee
where the terms have their usual meaning. We can now write $H_F$
explicitly in this case. The dispersion relation
(\ref{dispersion}) is actually a matrix equation and $H_F$ is the
matrix $E$ in the flavor basis. We assume that $M^2$ and $b$ are
diagonalised by the unitary matrices $U_m$ and $U_b$ respectively
so that \be H_F = U_m^\dagger diag(m_1^2,m_2^2) U_m + U_b^\dagger
diag(b_1,b_2) U_b \ee and the Eq.(\ref{eqofmot}) becomes \be
i\frac{d}{dt}{\pmatrix {\nue \cr \numu \cr}}= \left[\frac{\Delta
m^2}{4E}\pmatrix{-\cos2\theta_m & \sin2\theta_m \cr \sin2\theta_m
& \cos2\theta_m} + \frac{\Delta b}{2} \pmatrix{-\cos2\theta_b &
\sin2\theta_b \cr \sin2\theta_b & \cos2\theta_b} \right ]
{\pmatrix {\nue \cr \numu \cr}} \label{eq2} \ee where $\Delta m^2$
is the mass squared difference in vacuum and $\Delta b = b_2 -
b_1$ is the difference between the eigenvalues of the matrix $b$.
The mixing angles $\theta_m$ and $\theta_b$ correspond to the
rotation angles that diogonalises the mass matrix and $b$
respectively. Since there are two phases, corresponding to the two
mixing matrices, there will be an extra phase which cannot be
absorbed into the neutrino fields. But we have put that to zero 
for simplicity. It is straightforward to include it.

We are interested in the evolution of the neutrino states. Let us
define \bea a_{ee} &=& \langle \nue | \nue(t) \rangle
\\
a_{e\mu} &=& \langle \nue | \numu(t) \rangle \eea Then
Eq.(\ref{eq2}) could be written as \be i\frac{d}{dt}{\pmatrix
{a_{ee} \cr a_{e\mu} \cr}}= \left[\frac{\Delta
m^2}{4E}\pmatrix{-\cos2\theta_m & \sin2\theta_m \cr \sin2\theta_m
& \cos2\theta_m} + \frac{\Delta b}{2} \pmatrix{-\cos2\theta_b &
\sin2\theta_b \cr \sin2\theta_b & \cos2\theta_b} \right ]
{\pmatrix {a_{ee} \cr a_{e\mu} \cr}} \label{eq3} \ee That means we
have two coupled differential equations \bea
i\frac{d}{dt}a_{ee} &=& -Aa_{ee} + Ba_{e\mu}\\
i\frac{d}{dt}a_{e\mu} &=& Ba_{ee} + Aa_{e\mu} \eea where \bea
A=\frac{\Delta m^2}{4E}\cos2\theta_m + \frac{\Delta b}{2}\cos2\theta_b\\
B=\frac{\Delta m^2}{4E}\sin2\theta_m + \frac{\Delta
b}{2}\sin2\theta_b \eea 
It is easy to solve these coupled equations using
the boundary conditions
\bea
a_{ee}&=&1 ~~{\rm at ~~t=0}\\
i\frac{d}{dt}a_{ee} &=& -A ~~{\rm at ~~t=0}
\eea
We get
\bea a_{ee} =
\frac{1}{2}\left\{ \left (1 + \frac{A}{(A^2 + B^2)^{1/2}}\right )
e^{-i(A^2+B^2)^{1/2}t} + \left ( 1 - \frac{A}{(A^2 + B^2)^{1/2}}\right ) 
e^{i(A^2+B^2)^{1/2}t}\right\}
\eea
The survival probability is just the
modulus squared of the amplitude $a_{ee}$ 
\bea P_{ee} = \cos^2(A^2
+ B^2)^{1/2}t + \frac{A^2}{(A^2+B^2)^{1/2}} \sin^2(A^2+B^2)^{1/2}t
\eea If we define $2A = \Delta\cos 2\Theta$ and
$2B=\Delta \sin 2\Theta$ then we
get the result \cite{Barger:2000iv},
\be
P_{ee}=1-\sin^22\Theta \sin^2(\Delta L/4)
\label{prcpt}
\ee
where $\Delta = 4((A^2 + B^2)^{1/2}$.

\section*{Appendix B: The lepton number violating ${\cal L}_{LIV}$} 
In this appendix 
we consider the new LIV term in Eq.\ref{cpt} and look for the equation of 
motion for the neutrinos. 
The Lagrangian for this case is,
\be
{\cal L} = {\cal L}_{mass} + {\cal L}_{LIV}
\ee
where ${\cal L}_{mass}$ 
contains the usual mass terms for the light 
neutrinos and ${\cal L}_{LIV}$ corresponds to 
the LIV operator in Eq.\ref{cpt}
but rewritten in 4-component Majorana notation:
\be
{\cal L}_{LIV}=-h^{\alpha\beta}_{\mu\nu}(\overline{\Psi^C_M})_\alpha
(i\Sigma^{\mu\nu})(\Phi_M)_\beta
+ H.C.
\label{lagliv}
\ee
$\Sigma_{\mu\nu}=\frac{i}{2}[\gamma_\mu,\gamma_\nu]$ and 
$\Psi_M$ and $\Phi_M$ are 4-component Majorana neutrino fields.
The Hermitian conjugate term may be absorbed into a redefinition of
the coefficient as follows:
\be
{\cal L}_{LIV}=-{\cal H}^{ab}_{\mu\nu}(\overline{\Psi^C_M})_a 
(i\Sigma^{\mu\nu})(\Phi_M)_b
\label{lag4liv}
\ee
It is easy to see from Eq.\ref{lag4liv} that the coefficients 
${\cal H}^{ab}_{\mu\nu}$ are antisymmetric and hence CPT is 
conserved.

We can write this Lagrangian in 
the two-component notation:
\bea
\sigmn &=&
\frac{1}{4}\left [\sigma^\mu \bar\sigma^\nu 
- \sigma^\nu \bar\sigma^\mu \right ]\\
 \barsigmn&=&\frac{1}{4}\left [ \bar\sigma^\mu \sigma^\nu - 
\bar\sigma^\nu \sigma^\mu \right ]
\eea
\be
\Sigma_{\mu\nu}=\pmatrix{i\sigmn & 0 \cr 0 & i\barsigmn}
\ee
where $\sigma^\mu=(I_2,{\bf \sigma})$ and 
$\sigma_\mu = \bar\sigma^\mu = (I_2,-{\bf \sigma})$.
The 4-component Majorana spinor can be written in terms of two 2-component 
objects as
\be
\Psi_M = \pmatrix {\sia \cr \sidot}
\ee
where $\alpha=1,2$ and $\sia$ is a \lh 2-component neutrino, while 
$\sidot=-i\sig\psi^*_\alpha$ is the corresponding CP conjugated spinor field.
For Majorana spinors
\be
\Psi_M^C = \Psi_M
\ee
And
\bea
\overline{\Psi^C_M} &=& {\Psi_M^C}^\dagger \gamma_0\\
&=& \pmatrix {\overline\psi_{\dot\alpha} & \psi^\alpha} \gamma_0
\eea
For 
\be
\gamma_0 = \pmatrix{0&1\cr1&0}
\ee
we have
\be
\overline{\Psi^C_M} = \pmatrix{\psi^\alpha & \overline\psi_{\dot\alpha}}
\ee
Therefore the Lagrangian (\ref{lag4liv}) is given by
\bea
{\cal L}_{LIV}&=&
{\cal H}^{ab}_{\mu\nu}\left [ 
\pmatrix{\psi^\alpha & \overline\psi_{\dot\alpha}} 
\pmatrix{\sigmn & 0 \cr 0 & \barsigmn }
\pmatrix {\phi_\beta \cr \overline\phi^{\dot\beta}} 
\right ]
\nonumber\\
&=&
{\cal H}^{ab}_{\mu\nu}\left [
\psi^\alpha {(\sigmn )_\alpha}^\beta\phi_\beta +
\overline\psi_{\dot\alpha}
{(\barsigmn )^\alpha}_\beta \overline\phi^{\dot\beta}
\right ]
\label{lagf}
\eea
where we use $\Psi_M$ to denote the spinor with flavor $a$ and 
$\Phi_M$ to denote the spinor field of flavor $b$ and have suppressed 
the flavor index in the 2-component spinors.


Since ${\cal H}^{ab}_{\mu\nu}$ is antisymmetric 
we can express it in the Lorentz space as,
\bea
{\cal H}_{\mu\nu}=
\pmatrix{0 &-h_{01} &-h_{02} &-h_{03} \cr
h_{01} & 0 & -h_{12} & h_{13} \cr
h_{02} &h_{12} & 0 & -h_{23} \cr
h_{03} & -h_{13} &h_{23} & 0 }
\eea
where we have suppressed the flavor indices.
Then we have
\bea
\sigma^{\mu\nu}{\cal H}_{\mu\nu} &=&
\pmatrix{H_0 & H_+ \cr
            H_- & -H_0} 
\label{cpm0}
\eea
where
\bea
H_0 &=& h_{12} + h_{03} \\
H_\pm &=& (h_{23} + h_{01}) \pm i(h_{13}
+h_{02})
\eea

The first term in
Eq. (\ref{lagf}) can be seen as either two incoming left-handed neutrinos 
of different flavors or
alternatively as an incoming left-handed neutrino and an
out-going right-handed neutrino of a different flavor.
Therefore there is a flip of
flavor as well as spin in Eq. (\ref{lagf}).
In the ultra-relativistic limit
the full Lagrangian can be written as 
\be
{\cal L}_{LIV}=\bigg[(\nu^C_{eR})^\dagger \nu_{\mu L} H_+
-(\nu_{eL})^\dagger \nu_{\mu R}^C H_-
+(\nu_{\mu L})^\dagger \nu_{eR}^C H_- - (\nu_{\mu R}^C)^\dagger \nu_{eL}H_+
\bigg ]
\ee

We use this ${\cal L}_{LIV}$ to get the equation of motion 
for the neutrinos.


\end{document}